**Divergence preservation in the ADI algorithms for electromagnetics**

David N. Smithe, John R. Cary*, Johan A. Carlsson
Tech-X Corporation
5621 Arapahoe Ave., Suite A
Boulder, CO 80303

Email: smithe@txcorp.com
Phone: 303-996-2023
Fax: 303-448-7756

* Also: Dept. of Physics, University of Colorado, Boulder, CO 80309

## Abstract

The recent advances in alternating direct implicit (ADI) methods promise important new capability for time domain plasma simulations, namely the elimination of numerical stability limits on the time step. But the utility of these methods in simulations with charge and current sources, such as in electromagnetic particle-in-cell (EMPIC) computations, has been uncertain, as the methods introduced so far do not have the property of divergence preservation. This property is related to charge conservation and self-consistency, and is critical for accurate and robust EMPIC simulation. This paper contains a complete study of these ADI methods in the presence of charge and current sources. It is shown that there are four significantly distinct cases, with four more related by duality. Of those, only one preserves divergence and, thus, is guaranteed to be stable in the presence of moving charged particles. Computational verification of this property is accomplished by implementation in existing 3D-EMPIC simulation software. Of the other three cases, two are verified unstable, as expected, and one remains stable, despite the lack of divergence preservation. This other stable algorithm is shown to be related to the divergence preserving case by a similarity transformation, effectively providing the complement of the divergence preserving field in the finite-difference energy quantity.

## 1. Introduction

Electromagnetic simulations of plasma can be made more robust with the use of implicit solvers, since these permit use of time step consistent with the phenomena of interest, rather than the fastest time-scale of the plasma, e.g., the plasma frequency, or speed-of-light Courant limit. An interesting class of those solvers is that of so-called Alternating Direction Implicit (ADI) solvers, of which one example was introduced in [1] and further analysis was presented in [2]. These solvers are found by separating the time development operator (curl acting on **E** and **B**) for Maxwell's equations into two parts. In each part, the derivative in any one direction (say x) acts on only one of the pairs (say $E_y$ and $B_z$) of the combined six components of the electric and magnetic fields. This allows one to develop a full update through Strang [3] splitting, which, upon repetitive application, alternates the updates of each of the operator parts separately. Furthermore, the simplicity of each part allows fully implicit updates with the requirement of only one-dimensional, tridiagonal solves, which can be rapidly effected through the Thomas algorithm [4].

In this paper we investigate how these methods might be combined with self-consistent charges and current, with the particular example being electromagnetic particle-in-cell (EMPIC) methods [5,6]. For such situations, the importance of charge conservation has been noted. A charge conserving current deposition algorithm has the property that integration of the finite difference continuity equation, $\Delta\rho = -\nabla_{FD}\cdot\mathbf{j}\Delta t$ (ρ is the charge density, **j** is the current density, and $\nabla_{FD}$ is the finite-difference form of the divergence operator), gives a charge density that is consistent with the charge density of the locations of the particles in the simulations. Such a current deposition algorithm was developed by Villaseñor and Buneman [7] and later in [8] for smoother particles. As demonstrated in [9], failure to have a charge conserving current deposition algorithm can lead to nonphysical divergence buildup in the electric field that ultimately leads to catastrophic failure of the simulation.

The second issue, which is dealt with in this paper, is whether the electromagnetic update algorithm preserves this divergence of the current. Should indeed this be the case, then one has the highly desirable property that, if $\nabla_{FD}\cdot\mathbf{B} = 0$ and $\nabla_{FD}\cdot\mathbf{E} = \rho/\varepsilon_0$ at one time-step, then these remain true, to machine precision, after application of the update operator, to the next time step. Stated another way, for the magnetic update, the requirement is that a divergence-less field, upon update, leads to a divergence-less field. For the electric update, the associated condition is that, given a conservative charge deposition scheme, Gauss's Law, $\nabla_{FD}\cdot\Delta\mathbf{E} = -\nabla_{FD}\cdot\mathbf{j}\Delta t/\varepsilon_0$, remains valid, to machine precision, after application of the update operator, without the need for any additional divergence cleaning step. Such divergence preservation is indeed the case for the Yee [10] update and for a Crank-Nicholson update as well, but it is not *a priori* clear for the ADI updates.

In this paper, we perform the splitting of the curl operator in Maxwell's equations into two parts, denoted **M** and **P**, leading to the four fundamental operators of ADI algorithms, which we denote as $\left(1+\frac{\Delta t}{2}\mathbf{M}\right)$, $\left(1+\frac{\Delta t}{2}\mathbf{P}\right)$, $\left(1-\frac{\Delta t}{2}\mathbf{M}\right)^{-1}$ and $\left(1-\frac{\Delta t}{2}\mathbf{P}\right)^{-1}$. The first two operators are explicit, and the second two are implicit, since they are inverses. An ADI algorithm uses all four of these operators to construct the full update sequence. There are $24 = 4!$ possible combinations of the four operators. Of these 24, 12 are equivalent to each other by simple duality, that is, interchange of **M** and **P**. Of the remaining 12 combinations, 4 are not transformable to Strang splitting, because successive alternation between **M** and **P** operators prevents construction of unitary operators. Finally, of the remaining 8 combinations, only 4 are time-reversible, which is an important property of the Maxwell update and assures a minimum of 2$^{nd}$-order accuracy. Of those four combinations, one arises naturally from that introduced by [1] but does not preserve divergence. Of the other three combinations, only one is found to be divergence preserving. Strikingly, this one divergence preserving ADI composition has not been previously identified or studied. Finally, the remaining two compositions are related to the previous two by similarity transformation, and as previously indicated, are not divergence preserving.

We then proceed to examine the consequences of using any of these updates in EMPIC, having implemented them in the VORPAL [11] computational application. We show that, like the previous results of [9] for charge non-conservation, the lack of divergence preservation in two of the algorithms, including the one most closely related to that of [1], leads to catastrophic failure of the associated ADI-EMPIC simulations. We show that the new, divergence preserving algorithm does not have this catastrophic failure. Nor does the algorithm related to it by a similarity transformation.

These results can be applied to ADI algorithms in simulations of many phenomena for which one would like to use EMPIC, but for which the Courant limit is constraining. One example comes from the requirement to resolve the plasma Debye length in order to avoid self-heating [12]. This requirement in an explicit EM simulation then leads to a Courant time step that is $\Delta t \approx \lambda_D / c$. However, if light waves are not important, one really needs a time step that can be much larger, of order the time for a thermal electron to cross a cell, or $\Delta t \approx \lambda_D / v_e$, where $v_e$ is the electron thermal speed. Thus, use of explicit methods, as might be needed to capture magnetic effects, requires a much smaller time step and, hence, much greater computational effort. Another common situation involves the propagation of a bunched, or otherwise spatially varying charge profile beam through a tube or cavity. The beam ultimately comes to equilibrium, with its self electric and magnetic forces in balance with any external fields and the beam divergence. Now the time scale is very long, as one is computing an equilibrium, so the Courant stability condition is even more restrictive. In general, the Courant condition is constraining whenever one is dealing with systems where one must keep magnetic effects yet light waves are not important to the dynamics.

The organization of this paper is as follows. In the following section we review the ADI-EM algorithms and derive the four fundamental operators from which all possible update operators can be assembled. We then show, in Sec. 3, using time-reversal arguments, that there are two possible second-order update operators for vacuum electromagnetics, and even those are related by a similarity transformation. In Sec. 4, we introduce current sources. This breaks the equivalence, so that there are four update operators, consisting of two pairs, with each in a pair related to the other by a similarity transformation. We show that only one of the four operators is divergence preserving. In Sec. 5 we present numerical results for ADI-EMPIC. These results show that only the divergence preserving operator and its relative by similarity are stable; the other two algorithms lead to catastrophic failure in ADI-EMPIC simulations. In Sec. 6 we derive the steady state solution for these two operators and show that the one similar to the divergence preserving update operator has the usual finite-differenced steady state Maxwell equations. In Sec. 7 we derive an energy invariant for the divergence preserving operator and its relative by similarity, and we show how its energy differs from the standard divergence preservation of FDTD EM. In Sec. 8 we discuss the use of these algorithms in conjunction with complex boundaries. The last section contains a summary and some conclusions.

## 2. ADI EM algorithms

The ADI algorithms derive from the fact that Maxwell's equations,

$$\frac{\partial \mathbf{B}}{\partial t} = -\nabla \times \mathbf{E} \tag{1a}$$

and

$$\frac{\partial \mathbf{E}}{\partial t} = -\frac{\mathbf{j}}{\varepsilon_0} + c^2 \nabla \times \mathbf{B}\,, \tag{1b}$$

can be written in the form,

$$\frac{\partial \tilde{\mathbf{V}}}{\partial t} = \tilde{\mathbf{S}} + \left(\tilde{\mathbf{P}} + \tilde{\mathbf{M}}\right)\tilde{\mathbf{V}}\,, \tag{2}$$

where $\tilde{\mathbf{V}}$ is the six-component field, (**E**, c**B**), $\tilde{\mathbf{S}}$ is the six-component field ($-\mathbf{j}/\varepsilon_0$, 0), and the operators $\tilde{\mathbf{P}}$ and $\tilde{\mathbf{M}}$ are defined by

$$\tilde{\mathbf{P}} \cdot \tilde{\mathbf{V}} = \tilde{\mathbf{P}} \cdot \begin{bmatrix} E_x \\ E_y \\ E_z \\ cB_x \\ cB_y \\ cB_z \end{bmatrix} \equiv c \begin{bmatrix} \partial cB_z/\partial y \\ \partial cB_x/\partial z \\ \partial cB_y/\partial x \\ \partial E_y/\partial z \\ \partial E_z/\partial x \\ \partial E_x/\partial y \end{bmatrix} \quad \text{and} \quad \tilde{\mathbf{M}} \cdot \tilde{\mathbf{V}} \equiv \tilde{\mathbf{M}} \cdot \begin{bmatrix} E_x \\ E_y \\ E_z \\ cB_x \\ cB_y \\ cB_z \end{bmatrix} \equiv -c \begin{bmatrix} \partial cB_y/\partial z \\ \partial cB_z/\partial x \\ \partial cB_x/\partial y \\ \partial E_z/\partial y \\ \partial E_x/\partial z \\ \partial E_y/\partial x \end{bmatrix}. \tag{3}$$

The mnemonic is that the operator $\tilde{\mathbf{P}}$ has a plus sign on the right, while the operator $\tilde{\mathbf{M}}$ has the minus sign. As can be seen, the interchange of $\tilde{\mathbf{P}}$ and $\tilde{\mathbf{M}}$ operators is equivalent to an interchange of **E** and $c$**B**, together with sign (parity) exchange. In the remainder of this paper we will assume what might be called an "**M**-first" choice of ADI-duality, and simply state here that the analysis proceeds identically under the interchange of **P** and **M** operators.

For numerical differencing, the Yee layout of the fields is assumed. In this layout, the electric fields are centered at the edges of a cell (indexed by $i,j,k$), while the magnetic fields are centered at the faces of the cell. We adopt the following notation to indicate the transition from field representation to discrete spatial representation: the previous tilde quantities are fields and calculus-operators, whereas without the tildes, the quantities are arrays and finite-differencing matrices. Thus, the finite difference forms of the above operators are

$$\mathbf{P}\cdot\mathbf{V} \equiv c\begin{bmatrix} c\left(B_{z,i,j,k} - B_{z,i,j-1,k}\right)/\Delta y \\ c\left(B_{x,i,j,k} - B_{x,i,j,k-1}\right)/\Delta z \\ c\left(B_{y,i,j,k} - B_{y,i-1,j,k}\right)/\Delta x \\ \left(E_{y,i,j,k+1} - E_{y,i,j,k}\right)/\Delta z \\ \left(E_{z,i+1,j,k} - E_{z,i,j,k}\right)/\Delta x \\ \left(E_{x,i,j+1,k} - E_{x,i,j,k}\right)/\Delta y \end{bmatrix} \quad \text{and} \quad \mathbf{M}\cdot\mathbf{V} \equiv -c\begin{bmatrix} c\left(B_{y,i,j,k} - B_{y,i,j,k-1}\right)/\Delta z \\ c\left(B_{z,i,j,k} - B_{z,i-1,j,k}\right)/\Delta x \\ c\left(B_{x,i,j,k} - B_{x,i,j-1,k}\right)/\Delta y \\ \left(E_{z,i,j+1,k} - E_{z,i,j,k}\right)/\Delta y \\ \left(E_{x,i,j,k+1} - E_{x,i,j,k}\right)/\Delta z \\ \left(E_{y,i+1,j,k} - E_{y,i,j,k}\right)/\Delta x \end{bmatrix}, \tag{4}$$

where now $\mathbf{V}$ is the full array of all values for all components and all cells, and $\mathbf{P}$ and $\mathbf{M}$ are linear operators on that space. Thus, the finite differenced Maxwell equations, in the absence of sources, are

$$\frac{\partial \mathbf{V}}{\partial t} = \left(\mathbf{P} + \mathbf{M}\right)\cdot \mathbf{V} \tag{5}$$

An important property of the discretized Maxwell system of equations is that the matrix operators are anti-symmetric, that is, $\mathbf{M}=-\mathbf{M}^T$ and $\mathbf{P}=-\mathbf{P}^T$, where the superscript '$T$' denotes transpose.

With Strang splitting, we break the above equation into two, and we solve each one separately, giving two advance operators. If each operator is found to second-order accuracy (third-order error), the full update can be found by application of the square root of the first operator followed by the second operator and then the square root of the first operator. More definitively, we first consider the equation,

$$\frac{\partial \mathbf{X}}{\partial t} = \mathbf{P}\cdot\mathbf{X} \tag{6}$$

which becomes

$$\mathbf{X}^{n+1} - \mathbf{X}^{n} = \tfrac{\Delta t}{2}\mathbf{P}\cdot\left(\mathbf{X}^{n+1} + \mathbf{X}^{n}\right) \tag{7}$$

upon 2nd-order time-centered finite differencing, at discrete time-steps, $t_n = n\Delta t$ using the usual time-superscript notation, $\mathbf{X}^n \equiv \mathbf{X}(t_n)$. The update solution of this equation is, of course,

$$\mathbf{X}^{n+1} = \mathbf{T}_{\mathbf{p}}\cdot\mathbf{X}^{n} \equiv \left(1 - \tfrac{\Delta t}{2}\mathbf{P}\right)^{-1}\cdot\left(1 + \tfrac{\Delta t}{2}\mathbf{P}\right)\cdot\mathbf{X}^{n} \quad , \tag{8}$$

thus defining the unitary second-order accurate time-advance matrix operator, $\mathbf{T}_\mathbf{P}$, based upon the original $\mathbf{P}$ operator. Similarly,

$$\mathbf{T}_\mathbf{M} \equiv \left(1 - \tfrac{\Delta t}{2}\mathbf{M}\right)^{-1}\cdot\left(1 + \tfrac{\Delta t}{2}\mathbf{M}\right) \tag{9}$$

gives the second-order accurate time-advance based upon the $\mathbf{M}$ operator. The unitary property of these matrices-operators, that is, $\mathbf{T_P}^T\mathbf{T_P}=\mathbf{1}$ and similarly for $\mathbf{T_M}$, is a result of the anti-symmetric property of the original **P** and **M** matrix-operators, and the commutability of the matrix factors, $\left(1+\frac{\Delta t}{2}\mathbf{M}\right)^{-1}$ and $\left(1-\frac{\Delta t}{2}\mathbf{M}\right)$, and analogously for **P**. Consequently, following Strang splitting, the composite operator,

$$\mathbf{T}_{Strang} \equiv \mathbf{T}_{\mathbf{M}}^{1/2}\mathbf{T}_{\mathbf{P}}\mathbf{T}_{\mathbf{M}}^{1/2}, \tag{10}$$

is also second-order accurate.

This is not the update proposed by [1]. Instead, they proposed the ADI-update operator,

$$\mathbf{T}_1 \equiv \left(1-\tfrac{\Delta t}{2}\mathbf{M}\right)^{-1}\cdot\left(1+\tfrac{\Delta t}{2}\mathbf{P}\right)\cdot\left(1-\tfrac{\Delta t}{2}\mathbf{P}\right)^{-1}\cdot\left(1+\tfrac{\Delta t}{2}\mathbf{M}\right), \tag{11}$$

which, we show in the next section, is also second-order accurate, although this property was not fully advertised in [1]. The two operators are related by a similarity transformation.

$$\mathbf{T}_1 = \left(1-\tfrac{\Delta t^2}{4}\mathbf{M}^2\right)^{-1/2}\cdot\mathbf{T}_{Strang}\cdot\left(1-\tfrac{\Delta t^2}{4}\mathbf{M}^2\right)^{1/2}. \tag{12}$$

Thus, the spectrums of the two updates are identical, and the update of one corresponds to the update of the similarity transform applied to the state vector of the other. Lee and Fornberg [2] noted that the operator of [1] could be rewritten,

$$\mathbf{T}_{GSS} \equiv \left(1-\tfrac{\Delta t}{2}\mathbf{M}\right)^{-1}\cdot\left(1-\tfrac{\Delta t}{2}\mathbf{P}\right)^{-1}\cdot\left(1+\tfrac{\Delta t}{2}\mathbf{P}\right)\cdot\left(1+\tfrac{\Delta t}{2}\mathbf{M}\right), \tag{13}$$

by virtue of the commutability of the middle two operators in Eq. (11). Application of this operator in an update equation for **V** allows the inverses to be taken in turn, producing a traditional ADI algorithm,

$$\mathbf{V}^{n+1}-\mathbf{V}^{n} = \tfrac{\Delta t}{2}\left(\mathbf{M}+\mathbf{P}\right)\cdot\left(\mathbf{V}^{n+1}+\mathbf{V}^{n}\right)-\tfrac{(\Delta t)^2}{4}\mathbf{P}\cdot\mathbf{M}\cdot\left(\mathbf{V}^{n+1}-\mathbf{V}^{n}\right) \tag{14}$$

that differs from the second-order Crank-Nicholson operator by the last term that is $O(\Delta t^3)$. Hence, the update operator (13) is second-order accurate. The notation adopted for this operator, "*GSS*", refers to "Gradient Steady State," due to the fact that this update, Eq. (14), allows a steady-state condition, $\mathbf{V}^{n+1}=\mathbf{V}^{n}$, whenever $\mathbf{V}^{n}$ is a pure gradient, since the gradient is in the null-space of the curl operator, $\left(\mathbf{M}+\mathbf{P}\right)$, for Yee-cell finite-differencing. This important property will be explored in more detail in Section 6.

## 3. Second-order accurate ADI operators

Inspired by the results of Lee and Fornberg, we approach this from another direction. Namely, we consider any and all possible operators that can be constructed from a product of each of the four operators appearing in Eq. (11), but we restrict ourselves to operators that are time-reversible. Time reversibility guarantees second-order accuracy, because time reversibility guarantees that the first non-vanishing term in the power series is third order in the time difference, just like in the actual evolution, and so the first possible difference is $O(\Delta t^3)$.

First we must understand how time reversal acts on the operators of Eq. (11). Since time reversal is computing the final state in terms of the initial, the order of the operators must be reversed, and the inverses interchange with multiplication. For an operator to be time reversal symmetric, then its inverse followed by $\Delta t \to -\Delta t$ must give the same operator. We see that the operators (11) and (13) introduced in Refs. 1 both have time-reversal symmetry and are, therefore, second-order accurate.

We now enumerate the other operators that can be time-reversal symmetric. As noted previously, we restrict ourselves to one specific choice of the $\mathbf{P} \leftrightarrow \mathbf{M}$ interchange duality, namely that all sequences of the form of Equation (11) can be taken to begin with either $\left(1-\frac{\Delta t}{2}\mathbf{M}\right)^{-1}$ or $\left(1+\frac{\Delta t}{2}\mathbf{M}\right)$ and end with the other corresponding time reversed operator. Thus, we have the previously defined update-operator from [1], and the operator that results from exchanging its first and last terms,

$$\mathbf{T}_1 \equiv \left(1-\tfrac{\Delta t}{2}\mathbf{M}\right)^{-1} \cdot \left(1+\tfrac{\Delta t}{2}\mathbf{P}\right) \cdot \left(1-\tfrac{\Delta t}{2}\mathbf{P}\right)^{-1} \cdot \left(1+\tfrac{\Delta t}{2}\mathbf{M}\right), \tag{15a}$$

and

$$\mathbf{T}_2 \equiv \left(1+\tfrac{\Delta t}{2}\mathbf{M}\right) \cdot \left(1+\tfrac{\Delta t}{2}\mathbf{P}\right) \cdot \left(1-\tfrac{\Delta t}{2}\mathbf{P}\right)^{-1} \cdot \left(1-\tfrac{\Delta t}{2}\mathbf{M}\right)^{-1}. \tag{15b}$$

Each of these two update operators has a corresponding equivalent, in the absence of currents, with the center terms commuted, including the previously noted Eq. (13).

$$\mathbf{T}_{GSS} \equiv \left(1-\tfrac{\Delta t}{2}\mathbf{M}\right)^{-1} \cdot \left(1-\tfrac{\Delta t}{2}\mathbf{P}\right)^{-1} \cdot \left(1+\tfrac{\Delta t}{2}\mathbf{P}\right) \cdot \left(1+\tfrac{\Delta t}{2}\mathbf{M}\right) \tag{15c}$$

$$\mathbf{T}_{DP} \equiv \left(1+\tfrac{\Delta t}{2}\mathbf{M}\right) \cdot \left(1-\tfrac{\Delta t}{2}\mathbf{P}\right)^{-1} \cdot \left(1+\tfrac{\Delta t}{2}\mathbf{P}\right) \cdot \left(1-\tfrac{\Delta t}{2}\mathbf{M}\right)^{-1} \tag{15d}$$

The notation adopted for this last operator, "*DP*", refers to "Divergence Preserving." This previously unheralded update operator will be the subject of the next section and is indeed the impetus for this study.

These two pairs of operators are related by similarity transformation,

$$\mathbf{T}_2 = \left(1-\frac{\Delta t^2}{4}\mathbf{M}^2\right) \cdot \mathbf{T}_1 \cdot \left(1-\frac{\Delta t^2}{4}\mathbf{M}^2\right)^{-1} \quad \text{and} \quad \mathbf{T}_{DP} = \left(1-\frac{\Delta t^2}{4}\mathbf{M}^2\right) \cdot \mathbf{T}_{GSS} \cdot \left(1-\frac{\Delta t^2}{4}\mathbf{M}^2\right)^{-1}. \tag{16}$$

Thus, in the absence of currents, there is only one fundamental second-order accurate vacuum electromagnetics update operator, with all others related to it by duality, commutability, or similarity transformation. Because the operator appearing in Eq. (16) arises repeatedly, we denote it as **R**, and discuss it further. Note that, in the continuous field representation, $\widetilde{\mathbf{M}}^2$ is a component-wise diagonal operator, which can be represented in derivative and spectral form as,

$$\widetilde{\mathbf{M}}^2 \cdot \widetilde{\mathbf{V}} = c^2 \begin{bmatrix} \partial^2 E_x / \partial z^2 \\ \partial^2 E_y / \partial x^2 \\ \partial^2 E_z / \partial y^2 \\ \partial^2 cB_x / \partial y^2 \\ \partial^2 cB_y / \partial z^2 \\ \partial^2 cB_z / \partial x^2 \end{bmatrix} = -c^2 \begin{bmatrix} k_z^2 E_x \\ k_x^2 E_y \\ k_y^2 E_z \\ k_y^2 cB_x \\ k_z^2 cB_y \\ k_x^2 cB_z \end{bmatrix}, \tag{17}$$

so that

$$\widetilde{\mathbf{R}} \equiv 1 - \frac{\Delta t^2}{4} \widetilde{\mathbf{M}}^2 = \begin{bmatrix} 1+\widetilde{\sigma}_z^2 & 0 & 0 & 0 & 0 & 0 \\ 0 & 1+\widetilde{\sigma}_x^2 & 0 & 0 & 0 & 0 \\ 0 & 0 & 1+\widetilde{\sigma}_y^2 & 0 & 0 & 0 \\ 0 & 0 & 0 & 1+\widetilde{\sigma}_y^2 & 0 & 0 \\ 0 & 0 & 0 & 0 & 1+\widetilde{\sigma}_z^2 & 0 \\ 0 & 0 & 0 & 0 & 0 & 1+\widetilde{\sigma}_x^2 \end{bmatrix}, \tag{18}$$

where for continuum fields, the spectral-form coefficient is

$$\widetilde{\sigma}_i = k_i c \Delta t / 2 \tag{19}$$

For discrete representation, the matrices, $\mathbf{M}^2$ and $\mathbf{R}$ are the block tridiagonal matrices based upon the $2^{nd}$-deriviative finite-differences, and the spectral-form coefficient is

$$\sigma_i = \frac{c\Delta t}{\Delta x} \sin\left(\frac{k_i \Delta x_i}{2}\right). \tag{20}$$

The $\mathbf{R}$ matrix is positive definite. For well resolved variations, $k_i\Delta x_i$,<<π and $k_i c\Delta t$<<2, it is approximately unity to second order, and departs from unity for poorly resolved variations in proportion to the Courant ratio, $c\Delta t/\Delta x$. Finally, we note that $\mathbf{R}$ contains only the $\mathbf{M}$ operator and not the $\mathbf{P}$ operator. Thus, $\mathbf{R}$ serves as an indicator of the choice of the ADI duality.

## 4. Divergence preservation

Divergence preservation becomes an issue when current sources are added into Maxwell's equations. In EMPIC, the electric current, $\mathbf{j}$, is usually computed at times halfway between the discrete times at which the electric field is known. Consequently, charge, ρ, is computed at the same times as the electric field, enabling Gauss's Law, with its divergence operation, to be evaluated. For ADI, we would like to continue to add the current in a time-centered manner, and in the simplest way possible. Thus, let us assume a current source evaluated at the half time step, $\mathbf{S}^{n+½}$. Thus, we look at possible update algorithms where the current is added once, in the center of the operators (15), e.g.,

$$\mathbf{V}_1^{n+1} = \mathbf{U}_1\left(\mathbf{V}_1^n\right) \equiv \left(1 - \tfrac{\Delta t}{2}\mathbf{M}\right)^{-1} \cdot \left(1 + \tfrac{\Delta t}{2}\mathbf{P}\right) \cdot \left[\left(1 - \tfrac{\Delta t}{2}\mathbf{P}\right)^{-1} \cdot \left(1 + \tfrac{\Delta t}{2}\mathbf{M}\right) \cdot \mathbf{V}_1^n + \Delta t \mathbf{S}^{n+1/2}\right], \tag{21a}$$

$$\mathbf{V}_2^{n+1} = \mathbf{U}_2\left(\mathbf{V}_2^n\right) \equiv \left(1 + \tfrac{\Delta t}{2}\mathbf{M}\right) \cdot \left(1 + \tfrac{\Delta t}{2}\mathbf{P}\right) \cdot \left[\left(1 - \tfrac{\Delta t}{2}\mathbf{P}\right)^{-1} \cdot \left(1 - \tfrac{\Delta t}{2}\mathbf{M}\right)^{-1} \cdot \mathbf{V}_2^n + \Delta t \mathbf{S}^{n+1/2}\right], \tag{21b}$$

$$\mathbf{V}_{GSS}^{n+1} = \mathbf{U}_{GSS}\left(\mathbf{V}_{GSS}^n\right) \equiv \left(1 - \tfrac{\Delta t}{2}\mathbf{M}\right)^{-1} \cdot \left(1 - \tfrac{\Delta t}{2}\mathbf{P}\right)^{-1} \cdot \left[\left(1 + \tfrac{\Delta t}{2}\mathbf{P}\right) \cdot \left(1 + \tfrac{\Delta t}{2}\mathbf{M}\right) \cdot \mathbf{V}_{GSS}^n + \Delta t \mathbf{S}^{n+1/2}\right], \tag{21c}$$

and

$$\mathbf{V}_{DP}^{n+1} = \mathbf{U}_{DP}\left(\mathbf{V}_{DP}^n\right) \equiv \left(1 + \tfrac{\Delta t}{2}\mathbf{M}\right) \cdot \left(1 - \tfrac{\Delta t}{2}\mathbf{P}\right)^{-1} \cdot \left[\left(1 + \tfrac{\Delta t}{2}\mathbf{P}\right) \cdot \left(1 - \tfrac{\Delta t}{2}\mathbf{M}\right)^{-1} \cdot \mathbf{V}_{DP}^n + \Delta t \mathbf{S}^{n+1/2}\right]. \tag{21d}$$

With this introduction of a current source term, the updates that were equivalent by virtue of commutation are no longer equivalent, since the source is added between the two commuting terms. Hence, there are four distinct second-order updates when current is present, rather than two, plus four more that can be obtained by duality.

In the Yee algorithm, divergence preservation comes from the fact that, as in the vector identity, $\nabla\cdot\nabla\times\mathbf{A}=0$, the numerical curl operator, $(\mathbf{P}+\mathbf{M})$, is in the null-space of the numerical divergence, which we denote here by $\nabla_{FD}\cdot$, and note that this provides the separate curls on the $\mathbf{E}$ and $c\mathbf{B}$ parts of the field. Thus, for any field array, $\mathbf{X}$,

$$\nabla_{FD} \cdot (\mathbf{P} + \mathbf{M}) \cdot \mathbf{X} = 0\,, \tag{22}$$

We now analyze the divergence preservation properties of Eq. (21d), $\mathbf{U}_{DP}$, on a field $\mathbf{V}_{DP}$, for which purpose we rewrite as

$$\left(1 - \tfrac{\Delta t}{2}\mathbf{P}\right) \cdot \left(1 + \tfrac{\Delta t}{2}\mathbf{M}\right)^{-1} \cdot \mathbf{V}_{DP}^{n+1} = \left(1 + \tfrac{\Delta t}{2}\mathbf{P}\right) \cdot \left(1 - \tfrac{\Delta t}{2}\mathbf{M}\right)^{-1} \cdot \mathbf{V}_{DP}^n + \Delta t \mathbf{S}^{n+1/2}\,. \tag{23}$$

Divergence preservation follows from the following identities,

$$\begin{aligned} \left(1 - \tfrac{\Delta t}{2}\mathbf{P}\right) \cdot \left(1 + \tfrac{\Delta t}{2}\mathbf{M}\right)^{-1} &= \left(1 - \tfrac{\Delta t}{2}\mathbf{P} - \tfrac{\Delta t}{2}\mathbf{M} + \tfrac{\Delta t}{2}\mathbf{M}\right) \cdot \left(1 + \tfrac{\Delta t}{2}\mathbf{M}\right)^{-1} \\ &= 1 - \tfrac{\Delta t}{2}\left(\mathbf{P} + \mathbf{M}\right) \cdot \left(1 + \tfrac{\Delta t}{2}\mathbf{M}\right)^{-1} \end{aligned} \tag{24}$$

and, similarly,

$$\left(1 + \tfrac{\Delta t}{2}\mathbf{P}\right) \cdot \left(1 - \tfrac{\Delta t}{2}\mathbf{M}\right)^{-1} = 1 + \tfrac{\Delta t}{2}\left(\mathbf{P} + \mathbf{M}\right) \cdot \left(1 - \tfrac{\Delta t}{2}\mathbf{M}\right)^{-1}. \tag{25}$$

Plugging Eqs. (24-25) into Eq. (23), and collecting terms results in an analogue to Eq. (14), although it is not as immediately recognizable as an ADI update,

$$\mathbf{V}_{DP}^{n+1} - \mathbf{V}_{DP}^n = \tfrac{\Delta t}{2}\left(\mathbf{M} + \mathbf{P}\right) \cdot \left[\left(1 + \tfrac{\Delta t}{2}\mathbf{M}\right)^{-1} \cdot \mathbf{V}_{DP}^{n+1} + \left(1 - \tfrac{\Delta t}{2}\mathbf{M}\right)^{-1} \cdot \mathbf{V}_{DP}^n\right] + \Delta t \mathbf{S}^{n+1/2} \tag{26}$$

We take the divergence of both sides of Eq. (26), and note from Equation (22) that the divergence, $\nabla_{FD}\cdot$, operating on $(\mathbf{M}+\mathbf{P})$ vanishes, thus yielding,

$$\nabla_{FD} \cdot \mathbf{V}_{DP}^{n+1} - \nabla_{FD} \cdot \mathbf{V}_{DP}^{n} = \nabla_{FD} \cdot \Delta t \mathbf{S}^{n+1/2}, \tag{27}$$

which is precisely the divergence preservation property, including source current. Since there are no magnetic currents, this result guarantees that the numerical divergence of the magnetic field always vanishes if it did so initially. Similarly, it implies that, provided the current and charge density satisfy the numerical continuity equation, the numerical divergence of the electric field will always equal the charge density if it did so originally. That is,

$$\nabla_{FD\mathbf{E}} \cdot \mathbf{E}_{DP}^{n+1} - \nabla_{FD\mathbf{E}} \cdot \mathbf{E}_{DP}^{n} = -\nabla_{FD\mathbf{E}} \cdot \Delta t \mathbf{j}^{n+1/2} / \varepsilon_0 = \left(\rho^{n+1} - \rho^{n}\right) / \varepsilon_0 . \tag{28}$$

where $\nabla_{FD\mathbf{E}}\cdot$ is the part of the divergence matrix associated with the electric field. The above manipulations relied on the fact that the first applied operator was an inverse, e.g., implicit, while the second was explicit, so that one has the identity plus a term whose left-most term is precisely the curl operator. Remarkably, of all the updates in (21), only $\mathbf{U}_{DP}$, (21d) has this form. Hence, none of the other updates will preserve divergence (though the dual obtained by interchanging **P** and **M** will have the same divergence preservation). All four updates have been implemented in numerical software to confirm this unique divergence preserving property, and also to verify the lack of divergence preservation in the other three updates. This will be discussed in further detail in Section 5.

In the absence of current sources, similarity transformations by the matrix **R** related these update operators, 1↔2 and *GSS*↔*DP*. Thus, we expect that these pairs will have similar stability properties in the presence of currents and charges. In particular, while the electric and magnetic fields for the update $\mathbf{U}_{GSS}$ are not updated in a divergence preserving way, they are related by a transformation to electric and magnetic fields that are updated in a divergence preserving way, e.g., to the electric and magnetic fields for the $\mathbf{U}_{DP}$ update. More precisely, the *DP* and *GSS* solutions are related by

$$\mathbf{V}_{DP}^{n} = \mathbf{R} \cdot \mathbf{V}_{GSS}^{n} . \tag{29}$$

Now, since the $\mathbf{V}_{DP}$ solution is divergence preserving, it implies that the *GSS* update preserves not divergence of the field, but rather divergence of the field multiplied by the **R** matrix, e.g., denoting $\nabla_{FD\mathbf{E}}\cdot$ and $\mathbf{R}_{\mathbf{E}}$ to be those parts of the divergence and **R** matrices associated with the electric field, we have,

$$\nabla_{FD\mathbf{E}} \cdot \mathbf{R}_{\mathbf{E}} \cdot \mathbf{E}_{GSS}^{n+1} - \nabla_{FD} \cdot \mathbf{R}_{\mathbf{E}} \cdot \mathbf{E}_{GSS}^{n} = -\nabla_{FD} \cdot \Delta t \mathbf{j}^{n+1/2} / \varepsilon_0 = \left(\rho^{n+1} - \rho^{n}\right) / \varepsilon_0 , \tag{30}$$

This operator, $\nabla_{FD\mathbf{E}} \cdot \mathbf{R}_{\mathbf{E}} \cdot$, has a larger stencil than the $\nabla_{FD\mathbf{E}} \cdot$. Also, as the spectral representation of the **R** matrix, Eq. (18), enhances larger wavenumbers, it acts as an anti-smoothing operator. In addition, as this operation, $\nabla_{FD\mathbf{E}} \cdot \mathbf{R}_{\mathbf{E}} \cdot \mathbf{E}_{GSS}$, resembles the use of non-trivial dielectric, $\boldsymbol{\varepsilon}$, in Gauss's Law, $\nabla\cdot\boldsymbol{\varepsilon}\cdot\mathbf{E}$, it is natural to think of the **R** matrix as an additional numerical permittivity/permeability associated with the ADI finite-differencing algorithm, especially if one recalls that **R** departs from the identity in proportion to worsening resolution of the field variations, and that **R** is an indicator of the choice of ADI duality. Further exploration of the role of the **R** matrix is provided in Sec. 6 and 7.

## 5. Numerical results for self-consistent charges and currents

To determine the effect of divergence preservation we developed prototype implementations within the VORPAL [11] computational framework, which can be used for EMPIC. We chose parameters to match the previous [9] numerical results. Namely, our simulations were 2D inside a box, 1m on a side and with 20x20 cells. The simulation was initialized to have no particles or fields, but then 5A of 30 keV electrons were injected from the middle third of the left wall. The time step was chosen to be the Courant value, and all four algorithms in (21) were tested. The simulations were run for 80,000 time steps, which is slightly over 900 times of transit of the beam across the system.

Figure 1 shows the results for the x-y scatter plot of the beam after 725 transit times using the ADI update operator, $\mathbf{U}_1$, based upon [1], with currents added in the center. One can see that the beam is developing an unphysical divergence, much like that seen in [9], when a non-charge conserving current deposition algorithm was used. While this is for one instant in time, further integration shows that the situation gets worse. It is accompanied by a growth in the divergence error, $\nabla_{FD} \cdot \mathbf{E} - \rho / \varepsilon_0$, which vanishes initially, but by this time in the simulation has grown to peak values of $3\text{x}10^6$ V/m$^2$. By comparison, the charge density for the beam is about $7.5\text{x}10^{-8}$ C/m$^3$, or $\rho / \varepsilon_0 \approx 8.5e3$ V/m$^2$. Thus the error in the divergence, by this time, is orders of magnitude greater than the actual beam charge density.

In contrast, Fig. 2 shows the results for the x-y scatter plot of the beam after 725 transit times using the newly introduced divergence preserving ADI update operator, $\mathbf{U}_{DP}$. There is no sign of beam divergence developing, as expected. The behavior is reminiscent of the standard Yee update with charge conserving current deposition that was shown in [9]. For this case, the divergence error is around $3\text{x}10^{-9}$ V/m$^2$, or around round-off error when compared with $\rho / \varepsilon_0 \approx 8.5e3$ V/m$^2$.

As noted earlier, the field solutions from update operators $\mathbf{U}_1$ and $\mathbf{U}_2$ are related by simple transformation. Hence they are expected to have identical stability properties in the presence of particles. Indeed, simulations show this to be the case. Use of the operator $\mathbf{U}_2$ also leads to unphysical beam divergence after a few hundred transit times.

Similarly, the field solution using $\mathbf{U}_{GSS}$ is stable, as it is related to field solution of $\mathbf{U}_{DP}$ by a simple transformation, and simulations using $\mathbf{U}_{GSS}$ are seen to have the same basic properties as those using $\mathbf{U}_{DP}$, i.e., no nonphysical beam divergence. For these simulations using $\mathbf{U}_{GSS}$, we observe a divergence error that is on the order of 300 V/m$^2$, which is less than, but not negligible, compared with the beam value of $\rho / \varepsilon_0 \approx 8.5e3$ V/m$^2$. So even though the usual divergence is not preserved by the operator, $\mathbf{U}_{GSS}$, the fact that a related divergence quantity is preserved appears to be enough to assure stability.

## 6. Steady state solutions

Because the update $\mathbf{U}_{DP}$ preserves the divergence, we are guaranteed that a solution using this operator always has $\nabla_{FD} \cdot \mathbf{B} = 0$ and $\nabla_{FD} \cdot \mathbf{E} = \rho / \varepsilon_0$, in exact analogy with the continuous-field representation of Maxwell's equations. However, the traditional Yee-cell finite-difference electromagnetics has the further property of steady-state solutions satisfying $\nabla_{FD} \times \mathbf{E} = 0$ and $\nabla_{FD} \times \mathbf{B} = \mu_0 \mathbf{j}$, also in analogy with the Maxwell's equations. Here we investigate the steady state solutions for some of the previously defined ADI updates.

We are particularly interested in the steady-state solution of the divergence preserving update, Eq. (26). Setting $\mathbf{V}_{DP}^{n+1}, \mathbf{V}_{DP}^{n} \rightarrow \mathbf{V}_{DP}$ in this equation, and noting that

$$\left(1+\tfrac{\Delta t}{2}\mathbf{M}\right)^{-1} \cdot \mathbf{V}_{DP}^{n+1} + \left(1-\tfrac{\Delta t}{2}\mathbf{M}\right)^{-1} \cdot \mathbf{V}_{DP}^{n} \rightarrow \left[\left(1+\tfrac{\Delta t}{2}\mathbf{M}\right)^{-1} + \left(1-\tfrac{\Delta t}{2}\mathbf{M}\right)^{-1}\right] \cdot \mathbf{V}_{DP} = 2\mathbf{R}^{-1} \cdot \mathbf{V}_{DP} \tag{31}$$

we can write the steady-state solution of the divergence preserving update in the form,

$$-\left(\mathbf{P}+\mathbf{M}\right) \cdot \mathbf{R}^{-1} \cdot \mathbf{V}_{DP} = \mathbf{S}, \tag{32a}$$

Noting the previously established relationship between $\mathbf{V}_{DP}$ and $\mathbf{V}_{GSS}$, Equation (29), it follows immediately that

$$-\left(\mathbf{P}+\mathbf{M}\right) \cdot \mathbf{V}_{GSS} = \mathbf{S}, \tag{32b}$$

which is the analogue to the continuous-field Maxwell equations in steady-state. Thus, the fields found from using $\mathbf{U}_{DP}$ do not satisfy the normal steady-state equations, rather they are related by the **R** matrix to the fields, $\mathbf{V}_{GSS}$, that do. Because the **R** matrix enhances shorter wavelengths, this ADI solution will have shorter wavelengths enhanced. To summarize, the G*SS* solution has analogous Maxwell steady state behavior, but it has divergence error. In contrast, the *DP* solution has no divergence error, but it does not satisfy the usual finite difference Maxwell equations in steady state.

It should be noted that this study has investigated the ADI algorithms for Maxwell equations in vacuum. After demonstrating the properties of the *DP* and *GSS* solutions, Equations (28) and (32b), we would be remiss not to point out the obvious similarity between the vacuum ADI algorithm and the Maxwell equations in the presence of non-trivial materials. When materials are present, the non-trivial dielectric creates a role separation between electric field, **E**, and electric displacement, $\mathbf{D}=\boldsymbol{\varepsilon}\cdot\mathbf{E}$, and similarly with $\mathbf{B}=\boldsymbol{\mu}\cdot\mathbf{H}$. In this role separation, it is the fields **D** and **B** which occur in the divergence equations, and it is the fields **E** and **H** which are acted upon directly by the curl operator, and thus occur in the steady-state equations. The *DP* and *GSS* vacuum solutions have similar role separation, namely $\mathbf{V}_{DP}$ occurs in the divergence equation, and $\mathbf{V}_{GSS}$ occurs in the steady-state equations, and furthermore, as already mentioned, the **R** matrix connects $\mathbf{V}_{DP}$ to $\mathbf{V}_{GSS}$ exactly as $\boldsymbol{\varepsilon}$ connects the **D** to **E** and $\boldsymbol{\mu}$ connects **B** to **H**. Thus, despite the fact that we are talking specifically of vacuum field solutions, it is not a difficult observation to note that the *DP* solution behaves analogously to **D** and **B** in a material, while the *GSS* solution behaves analogously to **E** and **H** in a material. Whether this observation is a useful artifice remains to be seen. More to the point, we leave the investigation of the ADI algorithms in the presence of actual materials for later study.

## 7. Energy conservation

We consider only the *GSS* and *DP* updates, as those are the only cases that are stable in the presence of charged particles. In [2] it is shown that $\mathbf{T}_{GSS}$ has a positive energy quantity, which corresponds to

$$\mathbf{W} \equiv \mathbf{V}_{GSS} \cdot \mathbf{R} \cdot \mathbf{V}_{GSS} = \mathbf{V}_{DP} \cdot \mathbf{R}^{-1} \cdot \mathbf{V}_{DP} = \mathbf{V}_{GSS} \cdot \mathbf{V}_{DP}\,, \tag{33}$$

where the 2nd and 3rd definitions follow from Eq. (29). This quantity is conserved, as is the case in that analysis, in the absence of charges and currents. The goal of this section is to determine how this quantity evolves in the presence of charges and currents. We seek the change in the vacuum energy quantity (33), which can be written in the form,

$$\Delta\mathbf{W} = \mathbf{V}_{GSS}^{n+1} \cdot \mathbf{V}_{DP}^{n+1} - \mathbf{V}_{GSS}^{n} \cdot \mathbf{V}_{DP}^{n}\,. \tag{34}$$

This quantity arises naturally if one multiplies each side of Eq. (26) by the term following the curl, noting that due to the anti-symmetric property of the curl, that term vanishes, hence leaving just

$$\begin{aligned}\left(\mathbf{V}_{DP}^{n+1} - \mathbf{V}_{DP}^{n}\right)\cdot\left[\left(1+\tfrac{\Delta t}{2}\mathbf{M}\right)^{-1}\cdot\mathbf{V}_{DP}^{n+1} + \left(1-\tfrac{\Delta t}{2}\mathbf{M}\right)^{-1}\cdot\mathbf{V}_{DP}^{n}\right] &= \\ \Delta t\mathbf{S}^{n+1/2}\cdot\left[\left(1+\tfrac{\Delta t}{2}\mathbf{M}\right)^{-1}\cdot\mathbf{V}_{DP}^{n+1} + \left(1-\tfrac{\Delta t}{2}\mathbf{M}\right)^{-1}\cdot\mathbf{V}_{DP}^{n}\right]&\end{aligned} \tag{35}$$

Substituting from Eq. (26) for the left values of $\mathbf{V}_{DP}$ gives,

$$\begin{aligned}\left(\mathbf{V}_{GSS}^{n+1} - \mathbf{V}_{GSS}^{n}\right)\cdot\left[\left(1-\tfrac{\Delta t}{2}\mathbf{M}\right)\cdot\mathbf{V}_{DP}^{n+1} + \left(1+\tfrac{\Delta t}{2}\mathbf{M}\right)\cdot\mathbf{V}_{DP}^{n}\right] &= \\ \Delta t\mathbf{S}^{n+1/2}\cdot\left[\left(1+\tfrac{\Delta t}{2}\mathbf{M}\right)^{-1}\cdot\mathbf{V}_{DP}^{n+1} + \left(1-\tfrac{\Delta t}{2}\mathbf{M}\right)^{-1}\cdot\mathbf{V}_{DP}^{n}\right]&\end{aligned} \tag{36}$$

Finally, we note that by virtue of the anti-symmetric property of the matrix $\mathbf{M}\cdot\mathbf{R}$, the quantity $\left(\mathbf{V}_{GSS}^{n+1} - \mathbf{V}_{GSS}^{n}\right)\cdot\mathbf{M}\cdot\left(\mathbf{V}_{DP}^{n+1} - \mathbf{V}_{DP}^{n}\right)$ is zero, leaving just,

$$\begin{aligned}\Delta\mathbf{W} = \mathbf{V}_{GSS}^{n+1}\cdot\mathbf{V}_{DP}^{n+1} - \mathbf{V}_{GSS}^{n}\cdot\mathbf{V}_{DP}^{n} &= \Delta t\mathbf{S}^{n+1/2}\cdot\left[\left(1+\tfrac{\Delta t}{2}\mathbf{M}\right)^{-1}\cdot\mathbf{V}_{DP}^{n+1} + \left(1-\tfrac{\Delta t}{2}\mathbf{M}\right)^{-1}\cdot\mathbf{V}_{DP}^{n}\right] \\ &= \Delta t\mathbf{S}^{n+1/2}\cdot\left[\left(1-\tfrac{\Delta t}{2}\mathbf{M}\right)\cdot\mathbf{V}_{GSS}^{n+1} + \left(1+\tfrac{\Delta t}{2}\mathbf{M}\right)\cdot\mathbf{V}_{GSS}^{n}\right]\end{aligned}. \tag{37}$$

Equation (37) is similar to the expression for mechanical energy in the usual explicit finite-difference Maxwell equations, with the novel feature being the presence of the factors that make up the **R** matrix. We see from Eq. (33) that since the **R** matrix acts as an anti-smoothing operator, the *GSS* field must contain less short-wavelength field amplitude than the *DP* field. Similarly, the ADI form of the mechanical energy is smoothed if expressed in terms of the *DP* fields, and anti-smoothed if expressed in terms of the *GSS* fields. A full study of fluctuations in this system is a subject for future research. Here we note that if there is a fixed amount of energy in each Fourier mode, as one might expect from equipartition, the electric and magnetic fluctuations will be larger for the *DP* case, as the mode energy has a smaller factor in front of it.

## 8. Boundary conditions

Usage of these operators with boundary conditions is straightforward. Typically, two different types of boundary conditions are used. Stair-step boundary conditions are those for which a whole cell is either included or excluded from the simulation region. Cut-cell boundary conditions [13,14], are those for which the magnetic update is modified through use of Faraday's law for the fraction of the face that is within the simulation region.

For stair-step boundary conditions, with a cell either entirely included or excluded from the simulation region, the boundary ultimately consists of the faces of all included cells. Boundary conditions are applied by setting the values of the electric field on all cell edges on the boundary. The magnetic update equations are then valid for all interior faces as well as those on the boundary. The matrices **P** and **M** are then found by setting to zero any matrix elements operating on an electric field for which the corresponding edge is exterior – e.g., not even on the boundary. These terms give the time derivatives of the magnetic field. For the complement, one can use the regular magnetic update everywhere, provided it is followed by setting the values of the electric field on all boundary edges to the boundary values.

For the Dey-Mittra type cut-cell boundary conditions [13], which provide second-order accuracy in global quantities, again the electric update is unchanged, except that the electric field for any edge wholly outside the simulation region is not updated but remains at the boundary value. However, the magnetic updates are modified by changing the coefficient of the electric fields in **P** and **M** by multiplying them by the relative length of the edge within the simulation region divided by the area of the face of the magnetic field that is within the simulation region. For faces entirely outside the simulation region, the corresponding elements of the **P** and **M** matrices can be set to zero.

The Dey-Mittra boundary conditions are known to lead to a reduction in the maximum stable time step, when traditional leap-frog updating is used. The uniformly stable update algorithm [14] eliminates this reduction. We note here the important fact that use of the ADI methods, in lieu of leap-frog, also eliminates this time-step reduction, since the ADI methods are stable for arbitrary time step.

## 9. Summary and conclusions

An exhaustive study of ADI update operators for electromagnetics in the presence of charges and currents has been completed, and all possible 2nd-order ADI update operators having time-centered current addition at only one place have been identified. Of these only our newly introduced divergence-preserving $\mathbf{U}_{DP}$ update from (21d) and its dual (from interchanging **P** and **M**) are divergence preserving. We also demonstrated the curl-steady-state property of the ADI algorithm of [2], generalizing it to include current, and resulting in the $\mathbf{U}_{GSS}$ update from (21c), and demonstrate that a simple transformation connects the divergence-preserving and curl-steady-state field solutions. Upon implementation in EMPIC software, it was observed that only $\mathbf{U}_{DP}$ and $\mathbf{U}_{GSS}$ are suitable for use in particle simulations, as the others show long time divergence error growth that leads to unphysical behavior.

This work was supported by the Department of Energy under grants DE-FG02-07ER84732, DE-FG02-04ER41317, DE-FC02-07ER41499, and PPPL subcontract S-006288-F.

**Figure 1**

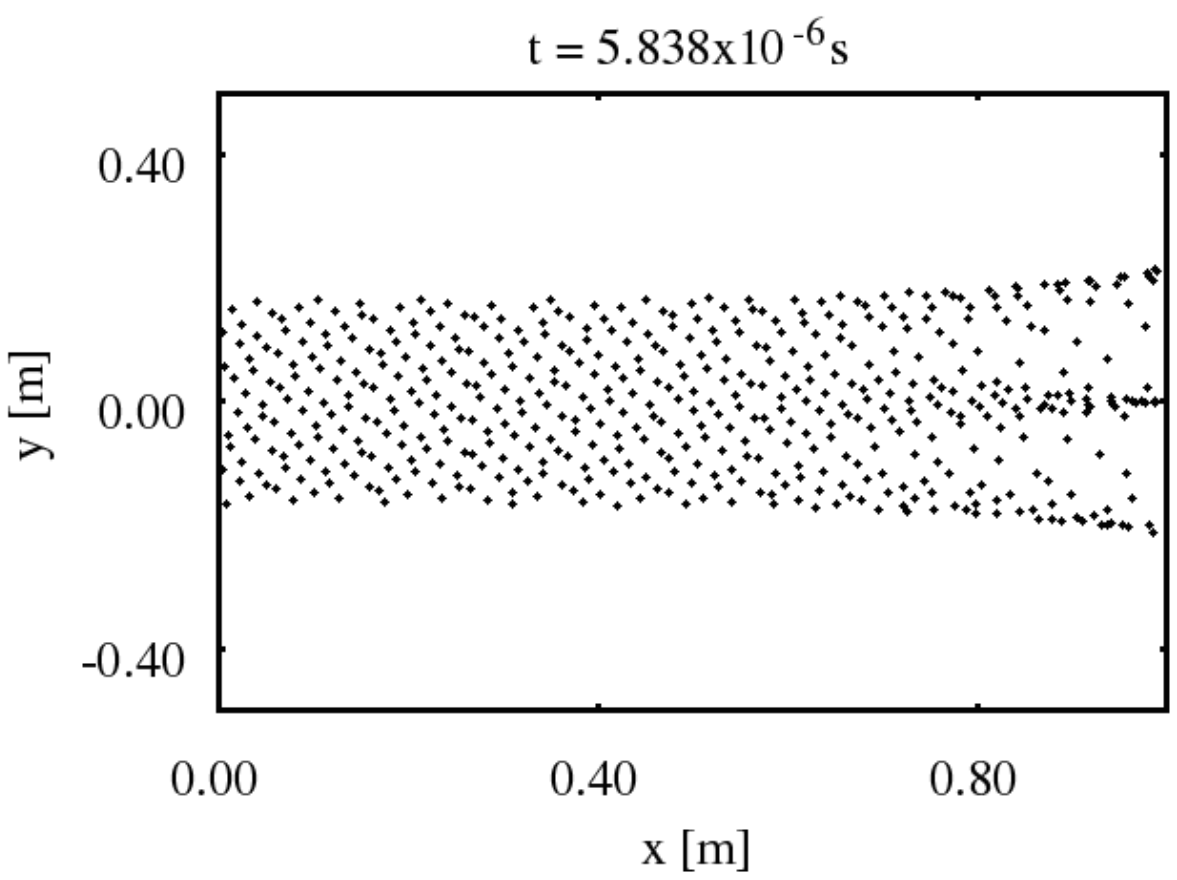

**Legend:** Configuration-space scatter plot of a beam after 575 transit times for the update operator $\mathbf{U}_1$. Artificial charge build-up on the grid causes an unphysical beam divergence instability.

**Submitted With:** Divergence preservation in the ADI algorithms for electromagnetics, by Smithe, Cary, and Carlsson

**Figure 2**

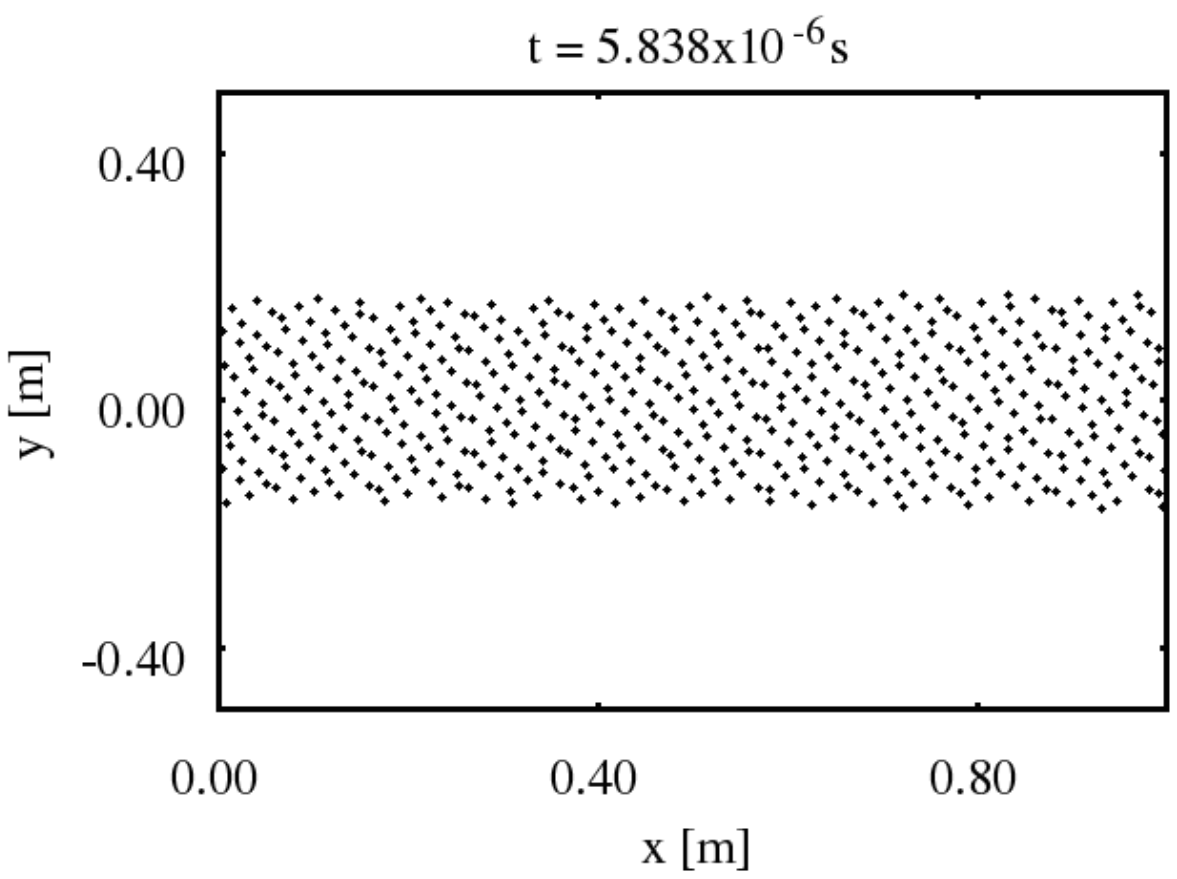

**Legend:** Beam after 775 transit times using the charge conserving update, $\mathbf{U}_{DP}$. Divergence error is zero to machine precision, and so beam transport is stable.

**Submitted With:** Divergence preservation in the ADI algorithms for electromagnetics, by Smithe, Cary, and Carlsson

**Figure Legends**

**Figure 1.** Configuration-space scatter plot of a beam after 575 transit times for the update operator $\mathbf{U}_1$. Artificial charge build-up on the grid causes an unphysical beam divergence instability.

**Figure 2.** Beam after 775 transit times using the charge conserving update, $\mathbf{U}_{DP}$. Divergence error is zero to machine precision, and so beam transport is stable.